\documentclass[9pt,twocolumn,twoside]{opticajnl}

\journal{ol} 

\setboolean{shortarticle}{true}

\title{Experimental upstream transmission of continuous variable quantum key distribution access network}

\author[1,*]{Xiangyu Wang}
\author[2,$\dagger$]{Ziyang Chen}
\author[1]{Zhenghua Li}
\author[1]{Dengke Qi}
\author[1]{Song Yu}
\author[2]{Hong Guo}

\affil[1]{State Key Laboratory of Information Photonics and Optical Communications, Beijing University of Posts and Telecommunications, Beijing 100876, China.}
\affil[2]{State Key Laboratory of Advanced Optical Communication Systems and Networks, School of Electronics and Center for Quantum Information Technology, Peking University, Beijing 100871, China.}

\affil[*]{Corresponding author: xywang@bupt.edu.cn, chenziyang@pku.edu.cn}

\begin{abstract}
Continuous-variable quantum key distribution which can be implemented using only low-cost and off-the-shelf components reveals great potential in the practical large-scale realization. Access network as a modern network necessity, connects multiple end-users to the network backbone. In this work, we demonstrate the first upstream transmission quantum access networks using continuous-variable quantum key distribution. A two-end-user quantum access network is then experimentally realized. Through phase compensation, data synchronization and other technical upgrades, we achieve 390kbps secret key rate of the total network. In addition, we extend the case of two-end-user quantum access network to the case of multiple users, and analyze the network capacity in the case of multiple users by measuring the additive excess noise from different time slots.
\end{abstract}

\setboolean{displaycopyright}{true}

\begin{document}

\maketitle
Quantum key distribution (QKD) which is designed to resist the malicious quantum computer~\cite{gisin2002quantum, pirandola2020advances} is important for future network security.
Continuous-variable (CV) QKD which encodes the key information into the quadrature components of the optical fields~\cite{weedbrook2012gaussian} is a strong candidate in realizing practical QKD systems~\cite{xu2020secure}.
Among which the GG02 protocol~\cite{grosshans2002continuous}, that can be implemented using only standard telecommunication devices, has thus gained a lot attentions. 
In recent years, many experimental demonstrations based on the coherent state and homodyne detection protocol have been conducted to show the utility~\cite{qi2007experimental, jouguet2013experimental, huang2016long, tian2022experimental, wang2022sub, chen2023continuous}.
With the rapid development of the CV-QKD realization, several networks have also been reported~\cite{aguado2019engineering,peev2009secoqc, huang2016field} to show the great possibility of building practical large-scale CV-QKD networks.

Yet, to develop a world-wide quantum internet~\cite{kimble2008quantum, pirandola2016physics}, diverse network structures are needed to fulfill the requirement of different concrete scenarios.
Access network as a special type of telecommunication network connects multiple subscribers with one backbone node, which is particularly suitable for the fiber-to-the-home scenario where internet services can be delivered to end-users. In a typical access network, optical network units (ONUs) are located at end-user sides, and the nodes that connects the network backbone are the optical line terminals (OLTs). The intermediate side between the ONUs and the OLT is the the optical distribution network (ODN). Signals in the access network can travel from the ONU to the OLT (in the upstream direction), or from the OLT to the ONU (in the downstream direction). More specifically, for an access network with upstream direction, signals from different ONUs are firstly combined at ONU, then forwarded to the OLT in a single fiber, which is the form of access network in our experiment. 

It is worth noting that CV-QKD has strong compatibility with existing communication systems, and has a high key rate within the distance of  metropolitan area network, which is very suitable for the implementation of access networks.
For CV-QKD, the quantum state can be deterministically distributed to the ONUs in the downstream direction. Although the quantum signal is broadcast to each ONU, it has been shown CV-QKD can be performed with the activated ONU in the downstream transmission scheme~\cite{Huang2021Realizing}. For practical implementations, it is proven that standard CV-QKD set-up is sufficient to be deployed in the access network of the downstream direction. On the other hand, the experimental demonstrations of CV-QKD access network with upstream direction are relatively less studied. At present, there is no demonstration of CV-QKD access network with upstream direction experimental system.

In this paper, we report the first experimental CV-QKD access network in the upstream direction. A 2-ONU access network is constructed, several techniques are developed to serve the access network. We achieve 390 kbps secret key rate of the total network, which includes 225 kbps of secret key rate of an ONU with transmission distance of 6 dB and 175 kbps of secret key rate of an ONU with transmission distance of 8 dB. In addition, We further explore the capacity of the access network, by assigning the siganls of one ONU into different time slots. This imitates a multi-ONU scenario, then by counting the additive excess noise in different spacing of the time slot, the result shows that up to 8 ONUs can be accessed to the network based on the current network implementation.
\begin{figure*}[htb]
\centerline{\includegraphics[width=14cm]{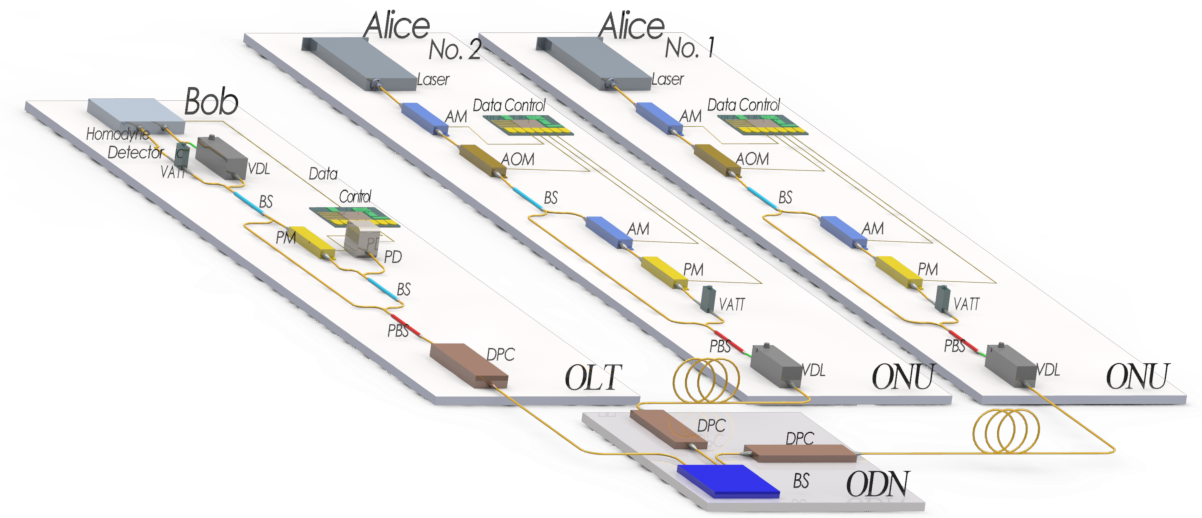}}
\caption{Schematic set-up of the two-end-user quantum upstream access network. Alice No. 1 and Alice No. 2 generate modulated coherent states (quantum signals) and local oscillators (LOs) which together compose the signals. The signals of each ONU then pass through a variable delay line (VDL) before being transmitted to the ODN. When the signals approach the ODN, dynamic polarization control (DPC) is firstly applied to perform the polarization compensation. The signals are then coupled to one stream by using the passive beamsplitter (BS). At the OLT, the combined signals are again being compensated for the polarization drifts by the DPC, then separated by the polarization beamsplitter (PBS) through polarization orientations. The LOs are then separated by two parts, where the majority of them interfere with the quantum signals. The rest of LOs are used to perform synchronizations and shot-noise unit (SNU) calibration. Lasers of Alice No. 1 and Alice No. 2 are of 1550.12 nm wavelength. AM: amplitude modulator; AOM: acousto-optical modulator; PM: phase modulator; VATT: variable attenuator; PD: photodiode.}
\end{figure*}
\begin{figure*}[htb]
\centerline{\includegraphics[width=17cm]{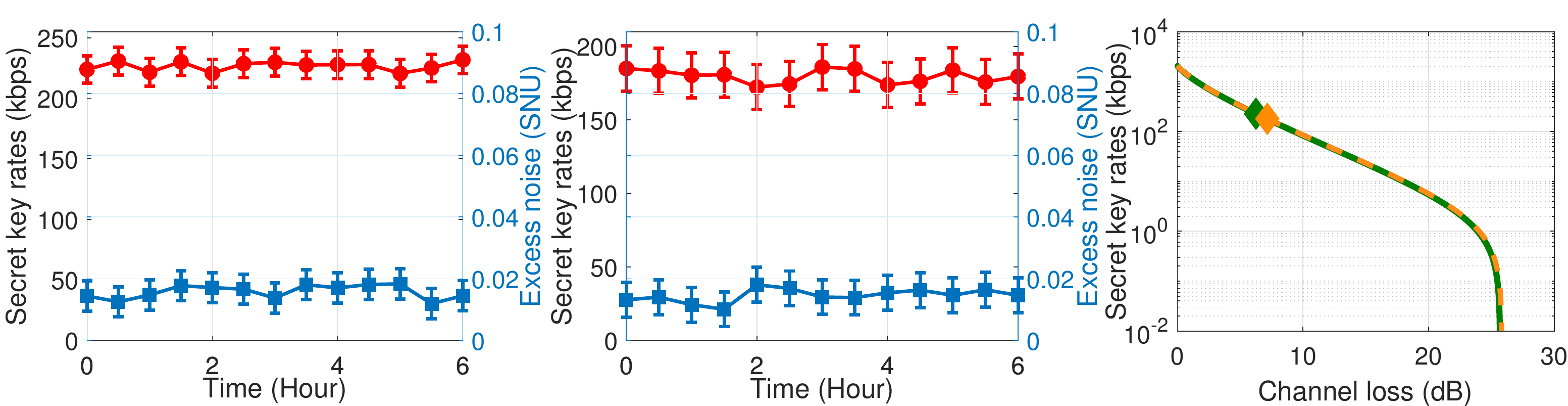}}
\caption{Secret key rates and excess noise (SNU) levels as a function of the experimental of (a) between ONU No. 1 and OLT, (b) between ONU No. 2 and OLT. Upper red marks are the secret key rates, lower blue marks represent the excess noise (SNU) levels. The modulation variance 4, and the reconciliation efficiency is around 0.95~\cite{wang2022continuous}. (c) Secret key rates curves of the experiment as a function of the channel loss, average secret key rate of 225 kbps of ONU No. 1 and Bob and average secret key rate of 175 kbps of ONU No. 2 and Bob are achieved under the finite-size regime.
}
\end{figure*}

The experimental set-up is depicted in Fig. 1, where we deployed a 2-ONU access network. The transmitters of CV-QKD (Alice No. 1 and Alice No. 2) are deployed at ONUs, the  receiver of CV-QKD (Bob) is placed at the OLT side.
The ONUs send quantum signals to the ODN through 10 km and 20 km standard single mode fibers individually. 
The passive 2:2 beam splitter (BS) is used as the passive optical network (PON) device in the ODN that couples the incoming signals from the ONUs (One port of the 2:2 BS is left unoccupied). The incoming signals time-multiplexed and alternatively gathered at the BS. The combined signals are then forwarded to the OLT through a single fiber with length of 5 km before being detected by the OLT. 

At ONUs’ side, The prepared quantum signals by the ONUs are basically the same for all Alice No.1  and Alice No. 2, where the Gaussian modulated coherent state protocol~\cite{grosshans2002continuous} is applied. Lasers produce continuous wave lights, which are then shaped by two cascaded amplititude modulators (AMs) into optical pulses of high extinction ratio. The ONU operates at a repetition frequency of 5 MHz and the duty circle of the pulses are all 10\%, which is 20 ns of time intervel. Pulses are then separated by high asymmetrical BS where the majority of them are used as local oscillators (LOs), the rest of the pulses are then modulated by a AM and a phase modulator (PM) according to the Gaussian distribution. The variable attenuator (VATT) subsequently weakens the quantum states to adjust the variance of the modulated quantum states to a desired level. The quantum states and the LOs are then recombined at the polarization beam splitter (PBS) in different polarization directions. To make sure the signals can arrive at the ODN in the corresponding time slots, extra patch cords are used to provide coarse adjustment, and the variable delay line (VDL) is applied for fine adjustment. This is to guarantee the precise allocation of quantum signals in the correct time slots. 

Before transmitting in a single fiber, the incoming signals are firstly being compensated. Because the quantum signals from each ONU go through different fibers, the polarization changes are not the same. The dynamic polarization control (DPC) module is firstly utilized to pre-compensate the polarization drifts from individual paths.
To achieve an upstream access network, it is critical the signals from each ONU should orderly fall into the corresponding time slot according to the time division multiplexing (TDM).
The CV-QKD transmitters operate at the repetition rate of 5 MHz, for a 3 ONUs scenario, the ODN forms time-slot of 50 ns, which is sufficiently available for both the quantum state and the LO to fit in. By adjusting the patch cords and the VDL, we make sure the quantum signals enter the ODN without any overlaps. The combined signals are then forwarded to the OLT through a single fiber.  

At OLT’s side, The optical layout for the OLT is quite the same as the standard CV-QKD receiver. When the signals arrive to the OLT Bob, another DPC is again applied to calibrated the polarization drifts, since the signals from each ONU travel through the same fiber, the polarization change is synergistic, so that a single DPC is sufficient to compensate the drifts.
The signals are then separated as the quantum signals and the LOs according to polarization orientations. The LOs are once more separated through a 90:10 BS, where the smaller portion of the LOs are then detected by the photodiode (PD) to provide synchronization information and phase compensation information.  
The rest of the LOs are then compensated by the PM, then interfere with the quantum states at the balanced BS. The outputs then pass through the VDL and the attenuator to equilibrize the common-mode signals. The outcomes are finally measured by homodyne detector.

To fulfill the access network requirement, several techniques have been upgraded. The synchronizations, sampling and data-processing have been upgraded to fulfill the access network requirement. For clock synchronization, the clock at the OLT is designed to serve as a master clock so that ONUs calibrate their own clocks according to the master clock. The data synchronization is performed individually with each ONU, the LOs that are used to perform data synchronization are specially modulated to specific levels, which separate them from each other. The sampling program is then able to recognize the signals from different ONUs. Subsequently in the data acquisition process, the programs independently sample the data of each ONU. Post-processing process like parameter estimation and security analysis are performed separately for each ONU as well. 

The experiment has collected ${5*{10^9}}$ data of completely 50 data set for each ONU, and the secret key rate results are shown in Fig. 2. In Fig. 2(a)and Fig. 2(b), we describe the changes in the secret  key rate and excess noise of ONU No. 1 and ONU No. 2 within 6 hours. Upper red marks are the secret key rates, lower blue marks represent the excess noise. The excess noise are all around 0.018(below 0.02). The secret key rate between ONU No. 1 and OLT is 200-250 Kbps and between ONU No. 1 and OLT is 150-200 Kbps. Furthermore, considering the finite-size, we give the relationship between the secret key rate and the channel loss, and mark the experimental results, as shown in Fig. 2(c).
The total loss between ONU No. 1 and OLT is around 6 dB, the obtained average secret key rate under the finite-size regime with block length of ${{\rm{1}}{{\rm{0}}^{{\rm{8}}}}}$ is about 225 kbps. For Alice No. 2 and Bob, the total loss is 8 dB, the averaged secret key rate under the finite-size regime with block length of ${{\rm{1}}{{\rm{0}}^{{\rm{8}}}}}$ is about 175 kbps.
The achieved secret key rates are quite high for CV-QKD systems of 5 MHz repetition frequency which exhibits great practicality of the CV-QKD access network.

Not restricted to the two-end-user access network, we also explore the maximum number of ONUs that are allowed to simultaneously access to the network using the same experimental implementations. However, with the increasing split ratio of the PON, the loss from the passive BS will inevitably increase. It also results in the increasing loss at the ODN, which will limit the performance of the access network. Thus, the BS-PON is replaced by the DWDM in further experiments. The DWDM-PON is more suitable to support more ONUs in the access network, since the insertion loss at each channel is rather fixed.

When more active ONUs are connected to the access network, the reserved time slot will ineluctably be shorten, the signals from each ONU tends to overlap with each other, which is a major obstacle in the TDM upstream scheme. Furthermore, the bandwidth of the homodyne detector also needs to increase to meet the increasing number of ONUs in  the network.
So, in this experiment, we narrow the signal pulses down to 10 ns, and another homodyne detector with the bandwidth of 100MHz is applied at the OLT.
Since the quantum signals and LOs are also multiplexed using TDM, the maximum time slot is about 25 ns. And that corresponds to at most 8 ONUs to access to the network at the same time if the repetition rate of different ONUs are limited at 5 MHz.

To show the feasibility of the multi-user quantum access network, the signals of Alice No. 2 are assigned to different time slots.
Since the averaging photon number of the quantum signals is very small which is practically negligible. In this experiment, Alice No. 2 only transmits LO pulses, while leaving the quantum signals unmodulated.
The additional excess noise is obtained by recording the excess noise value with and without the present of the LOs from Alice No. 2. The LOs from Alice No. 2 are assigned to the 4 closest time slots to the signals from Alice No. 1, the results are further normalized by the SNU. Next, we explain why the additive excess noise is only measured for the 4 closest time slots. For the current time slot, the signal is more likely to be deteriorated by the signals from its closest time slots which means the time slots that are just before and after the current time slot affect the current signal most. What is more, the damage caused to the previous and the next time slots can be viewed as the same because the signals are allocated to the center of the time slot. Therefore, by measuring the introduced excess noise from the 4 closest time slots, it is sufficient to evaluate excess noises introduced for the entire signal period of Alice No. 1.

\begin{figure*}[htb]
\centerline{\includegraphics[width=17cm]{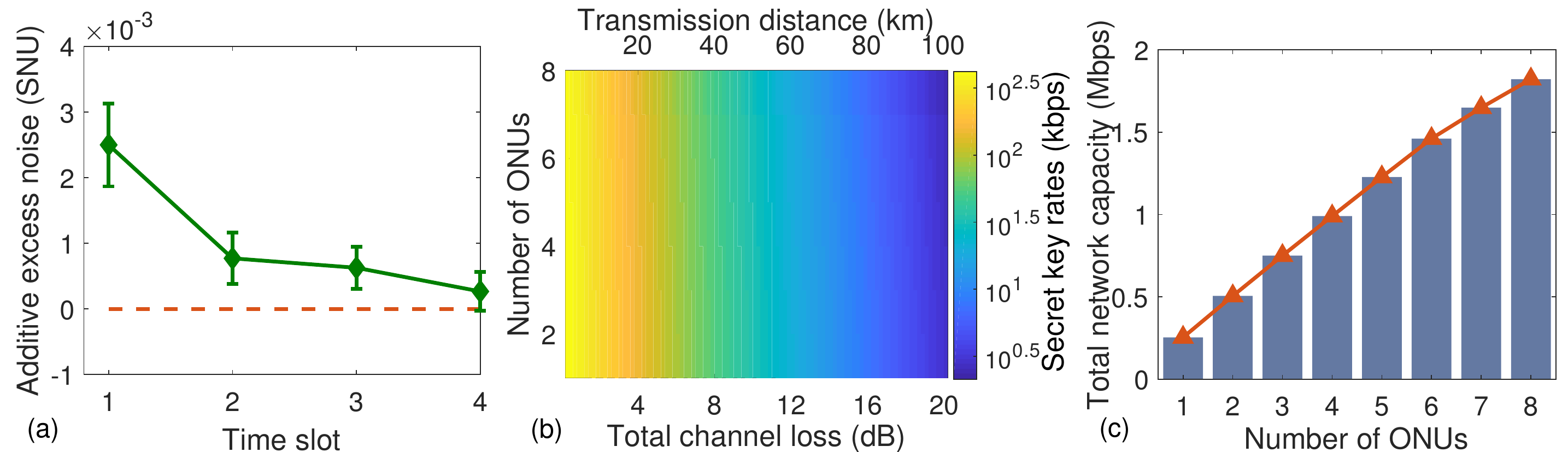}}
\caption{Network capacity analysis of the upstream access network. (a) Additive excess noise (SNU) as a function of different time slots. The experiment is conducted by assigning the LOs of Alice No. 2 to the 4 closest time slots compared to the signals of Alice No. 1. The time segregation between the signals of Alice No. 1 and the LOs of Alice No. 2 are 25 ns, 50 ns, 75 ns and 100 ns respectively. (b) Achievable secret key rates of a single ONU in the upstream access network.  The secret key rates are represented by pseudo-colors given by the number of ONUs that access to the network, and the total loss. In the simulation we consider the case where the recent joined ONU is assigned with the specified time slot that is as far away from the occupied time slots as possible. (c) Total network capacity as a function of the number of ONUs that access to the network. The signals from each ONU are assumed to endure the same total loss in the network.
}
\end{figure*}

The additive excess noise (SNU) as a function of the time slots is shown in Fig. 3(a). From the experiment, we do not observe too much increased excess noise as the signals of Alice No. 2 approach to the time slot that belongs to Alice No. 1. Also, the standard deviations are not increased much for the cases where the LOs are assigned to the time slots of 2, 3 and 4. The additive excess noise at time slot 1 is higher than the other cases, but is still within an acceptable range.

Based on the experimental results obtained above, the performance of the ONU in the access network is able to be simulated. In Fig. 3(b), achievable secret key rates of a single ONU within the upstream access network scheme given by the number of active ONUs and the total loss from the ONU to the OLT are provided.
The secret key rates which are represented by pseudo-colors do not exhibit significant dependence on the number of users access to the network which implies that the influence of the additive excess noise is rather subtle. It also suggests that 8 ONUs can operate in the upstream manner at the same time.
Thus, when the bandwidth of the homodyne detector is sufficiently high, by calibrating the signals from different ONUs into the pre-allocated time slots, the current access network can support support up to 8 active ONUs.

The above results also allow us to explore the secret key rate capacity of the entire upstream access network. The simulation result is shown in Fig. 3(c), where the total loss between each ONUs and the OLT is assumed to be the same as 6.26 dB. The bars are applied to indicate the network capacity given the number of ONUs that access to the network. It can be observed that the network capacity increases with the increasing number of ONUs connected to the network. Even so, as the number of ONUs increases, the increment in the network capacity is slightly slowing down. This is because the excess noise is also increased with the number of ONUs access to the network which results in the decreasing of the secret key rate of the individual ONU.

In conclusion, We demonstrated the first experimental system of  upstream transmission of CV-QKD access network in the case of standard optical fiber. In order to realize this system, we have developed new technologies such as phase compensation, data synchronization and clock synchronization, and finally realized 225 Kbps with 6 dB attenuation and 175 Kbps with 8 dB attenuation for two users. In addition, we explored the performance of the upstream transmission of CV-QKD access network when there are multiple users. With the increase of users, the total excess noise will increase, and the increase of the secret key rate will become slow.

\begin{backmatter}
\bmsection{Acknowledgments} This work was supported by the National Natural Science Foundation of China under Grant No. 62001041, No.62201012, the Fundamental Research Funds of BUPT under Grant No. 2022RC08, and the Fund of State Key Laboratory of Information Photonics and Optical Communications under Grant No. IPOC2022ZT09.

\bmsection{Disclosures} The authors declare no conflicts of interest.

\bmsection{Data availability} Data underlying the results presented in this paper are not publicly available at this time but may be obtained from the authors upon reasonable request.

\end{backmatter}

\bibliography{sample}

\begin{thebibliography}{10}
\newcommand{\enquote}[1]{``#1''}

\bibitem{gisin2002quantum}
N.~Gisin, G.~Ribordy, W.~Tittel, and H.~Zbinden, {\protect\JournalTitle{Reviews
  of Modern Physics}} \textbf{74}, 145 (2002).

\bibitem{pirandola2020advances}
S.~Pirandola, U.~L. Andersen, L.~Banchi, M.~Berta, D.~Bunandar, R.~Colbeck,
  D.~Englund, T.~Gehring, C.~Lupo, C.~Ottaviani \emph{et~al.},
  {\protect\JournalTitle{Advances in Optics and Photonics}} \textbf{12}, 1012
  (2020).

\bibitem{weedbrook2012gaussian}
C.~Weedbrook, S.~Pirandola, R.~Garc{\'\i}a-Patr{\'o}n, N.~J. Cerf, T.~C. Ralph,
  J.~H. Shapiro, and S.~Lloyd, {\protect\JournalTitle{Reviews of Modern
  Physics}} \textbf{84}, 621 (2012).

\bibitem{xu2020secure}
F.~Xu, X.~Ma, Q.~Zhang, H.-K. Lo, and J.-W. Pan, {\protect\JournalTitle{Reviews
  of Modern Physics}} \textbf{92}, 025002 (2020).

\bibitem{grosshans2002continuous}
F.~Grosshans and P.~Grangier, {\protect\JournalTitle{Physical Review Letters}}
  \textbf{88}, 057902 (2002).

\bibitem{qi2007experimental}
B.~Qi, L.-L. Huang, L.~Qian, and H.-K. Lo, {\protect\JournalTitle{Physical
  Review A}} \textbf{76}, 052323 (2007).

\bibitem{jouguet2013experimental}
P.~Jouguet, S.~Kunz-Jacques, A.~Leverrier, P.~Grangier, and E.~Diamanti,
  {\protect\JournalTitle{Nature Photonics}} \textbf{7}, 378 (2013).

\bibitem{huang2016long}
D.~Huang, P.~Huang, D.~Lin, and G.~Zeng, {\protect\JournalTitle{Scientific
  Reports}} \textbf{6}, 1 (2016).

\bibitem{tian2022experimental}
Y.~Tian, P.~Wang, J.~Liu, S.~Du, W.~Liu, Z.~Lu, X.~Wang, and Y.~Li,
  {\protect\JournalTitle{Optica}} \textbf{9}, 492 (2022).

\bibitem{wang2022sub}
H.~Wang, Y.~Li, Y.~Pi, Y.~Pan, Y.~Shao, L.~Ma, Y.~Zhang, J.~Yang, T.~Zhang,
  W.~Huang \emph{et~al.}, {\protect\JournalTitle{Communications Physics}}
  \textbf{5}, 1 (2022).

\bibitem{chen2023continuous}
Z.~Chen, X.~Wang, S.~Yu, Z.~Li, and H.~Guo, {\protect\JournalTitle{npj Quantum
  Information}} \textbf{9}, 28 (2023).

\bibitem{aguado2019engineering}
A.~Aguado, V.~Lopez, D.~Lopez, M.~Peev, A.~Poppe, A.~Pastor, J.~Folgueira, and
  V.~Martin, {\protect\JournalTitle{IEEE Communications Magazine}} \textbf{57},
  20 (2019).

\bibitem{peev2009secoqc}
M.~Peev, C.~Pacher, R.~All{\'e}aume, C.~Barreiro, J.~Bouda, W.~Boxleitner,
  T.~Debuisschert, E.~Diamanti, M.~Dianati, J.~Dynes \emph{et~al.},
  {\protect\JournalTitle{New Journal of Physics}} \textbf{11}, 075001 (2009).

\bibitem{huang2016field}
D.~Huang, P.~Huang, H.~Li, T.~Wang, Y.~Zhou, and G.~Zeng,
  {\protect\JournalTitle{Optics Letters}} \textbf{41}, 3511 (2016).

\bibitem{kimble2008quantum}
H.~J. Kimble, {\protect\JournalTitle{Nature}} \textbf{453}, 1023 (2008).

\bibitem{pirandola2016physics}
S.~Pirandola and S.~L. Braunstein, {\protect\JournalTitle{Nature}}
  \textbf{532}, 169 (2016).

\bibitem{Huang2021Realizing}
Y.~Huang, T.~Shen, X.~Wang, Z.~Chen, B.~Xu, S.~Yu, and H.~Guo,
  {\protect\JournalTitle{Physical Review Applied}} \textbf{16}, 064051 (2021).

\bibitem{wang2022continuous}
X.~Wang, H.~Wang, C.~Zhou, Z.~Chen, S.~Yu, and H.~Guo,
  {\protect\JournalTitle{Optics Express}} \textbf{30}, 30455 (2022).

\end{thebibliography}


\end{document}